\documentclass[twocolumn]{aastex61}
\usepackage{url}
\usepackage{ulem}

\newcommand{\appropto}{\mathrel{\vcenter{\offinterlineskip\halign{\hfil$##$\cr\propto\cr\noalign{\kern2pt}\sim\cr\noalign{\kern-2pt}}}}}
\newcommand{\bjdtdb}{\ensuremath{\rm {BJD_{TDB}}}}
\newcommand{\feh}{\ensuremath{\left[{\rm Fe}/{\rm H}\right]}}
\newcommand{\teff}{\ensuremath{T_{\rm eff}}}

\newcommand{\msun}{\ensuremath{\,M_\Sun}}
\newcommand{\rsun}{\ensuremath{\,R_\Sun}}
\newcommand{\lsun}{\ensuremath{\,L_\Sun}}
\newcommand{\mj}{\ensuremath{\,M_{\rm J}}}
\newcommand{\rj}{\ensuremath{\,R_{\rm J}}}
\newcommand{\fave}{\langle F \rangle}
\newcommand{\fluxcgs}{10$^9$ erg s$^{-1}$ cm$^{-2}$}
\begin{document}

\title{Transiting Exoplanet Monitoring Project (TEMP). IV. Refined System Parameters, Transit Timing Variations and Orbital Stability of the Transiting Planetary System HAT-P-25}

\correspondingauthor{Songhu Wang}
\email{song-hu.wang@yale.edu}

\author{Xian-Yu Wang}
\affiliation{Shandong Provincial Key Laboratory of Optical Astronomy and Solar-Terrestrial Environment, Institute of Space Sciences, Shandong University, Weihai 264209, China}

\author{Songhu Wang}
\affiliation{Department of Astronomy, Yale University, New Haven, CT 06511, USA}
\affiliation{\textit{51 Pegasi b} Fellow}

\author{Tobias C. Hinse}
\affiliation{Korea Astronomy and Space Science Institute, Daejeon 305-348, Republic of Korea}

\author{Kai Li}
\affiliation{Shandong Provincial Key Laboratory of Optical Astronomy and Solar-Terrestrial Environment, Institute of Space Sciences, Shandong University, Weihai 264209, China}

\author{Yong-Hao Wang}
\affiliation{Key Laboratory of Optical Astronomy, National Astronomical Observatories, Chinese Academy of Sciences, Beijing 100012, China}
\affiliation{School of Astronomy and Space Science, University of Chinese Academy of Sciences, Beijing 101408, China}

\author{Gregory Laughlin}
\affiliation{Department of Astronomy, Yale University, New Haven, CT 06511, USA}

\author{Hui-Gen Liu}
\affiliation{School of Astronomy and Space Science and Key Laboratory of Modern Astronomy and Astrophysics in Ministry of Education, Nanjing University, Nanjing 210093, China}

\author{Hui Zhang}
\affiliation{School of Astronomy and Space Science and Key Laboratory of Modern Astronomy and Astrophysics in Ministry of Education, Nanjing University, Nanjing 210093, China}

\author{Zhen-Yu Wu}
\affiliation{Key Laboratory of Optical Astronomy, National Astronomical Observatories, Chinese Academy of Sciences, Beijing 100012, China}
\affiliation{School of Astronomy and Space Science, University of Chinese Academy of Sciences, Beijing 101408, China}

\author{Xu Zhou}
\affiliation{Key Laboratory of Optical Astronomy, National Astronomical Observatories, Chinese Academy of Sciences, Beijing 100012, China}
\affiliation{School of Astronomy and Space Science, University of Chinese Academy of Sciences, Beijing 101408, China}

\author{Ji-Lin Zhou}
\affiliation{School of Astronomy and Space Science and Key Laboratory of Modern Astronomy and Astrophysics in Ministry of Education, Nanjing University, Nanjing 210093, China}

\author{Shao-Ming Hu}
\affiliation{Shandong Provincial Key Laboratory of Optical Astronomy and Solar-Terrestrial Environment, Institute of Space Sciences, Shandong University, Weihai 264209, China}

\author{Dong-Hong Wu}
\affiliation{School of Astronomy and Space Science and Key Laboratory of Modern Astronomy and Astrophysics in Ministry of Education, Nanjing University, Nanjing 210093, China}

\author{Xi-Yan Peng}
\affiliation{Key Laboratory of Optical Astronomy, National Astronomical Observatories, Chinese Academy of Sciences, Beijing 100012, China}
\affiliation{School of Astronomy and Space Science, University of Chinese Academy of Sciences, Beijing 101408, China}

\author{Yuan-Yuan Chen}
\affiliation{Purple Mountain Observatory, Chinese Academy of Sciences, Nanjing 210008, China}

\begin{abstract}

We present eight new light curves of the transiting extra-solar planet HAT-P-25b obtained from 2013 to 2016 with three telescopes at two observatories. We use the new light curves, along with recent literature material, to estimate the physical and orbital parameters of the transiting planet. Specifically, we determine the mid-transit times (T$_{C}$) and update the linear ephemeris, T$_{C[0]}$=2456418.80996$\pm$0.00025 [$\mathrm{BJD}_\mathrm{TDB}$]
 and P=3.65281572$\pm$0.00000095 days. We carry out a search for transit timing variations (TTVs), and find no significant TTV signal at the $\Delta T=$80 s-level, placing a limit on the possible strength of planet-planet interactions ($\mathrm{TTV_{G}}$). In the course of our analysis, we calculate the upper mass-limits of the potential nearby perturbers. Near the 1:2, 2:1, and 3:1 resonances with HAT-P-25b, perturbers with masses greater than 0.5, 0.3, and 0.5 $\mathrm{M_{\earth}}$ respectively, can be excluded. Furthermore, based on the analysis of TTVs caused by light travel time effect (LTTE) we also eliminate the possibility that a long-period  perturber exists with $M_{\rm p}> 3000 \,\mathrm{M_{J}}$ within $a=11.2\,{\rm AU}$ of the parent star.

\end{abstract}


\section{Introduction}

Thousands of transiting exoplanets have opened up a wealth of opportunities to discern nuances of the planetary formation and evolution processes.

In favorable cases, the measurement of planetary radii using transit photometry, combined with follow-up Doppler velocimetery (RV) measurements to determine masses, can reveal the bulk densities of representatives from the host of newly discovered super-Earths, determining, in turn, whether they are likely to be predominantly gaseous, watery or rocky worlds.
Moreover, RV observations taken during transit permit measurement of projected spin-orbital misalignment angles, $\lambda$, of the transiting planets (e.g., \citealt{2000A&A...359L..13Q}; \citealt{2005ApJ...631.1215W}; \citealt{2014arXiv1403.0652A}; \citealt{2015ARA&A..53..409W}; \citealt{wang2018}).

High-precision photometric follow-up observations not only can confirm the planetary interpretation of a transit-signal detection, but they can also contribute to improve the accuracy of the planet's physical and orbital parameters \citep{2017arXiv171206297W}. Moreover, high-precision photometric follow-up enables TTV assessments (e.g., \citealt{holman2005}; \citealt{agol2005}), which offer the prospect of detecting dynamically interesting perturbers. The architectures of systems that contain hot Jupiters provide clues that can potentially distinguish between competing formation theories for hot Jupiters \citep{Batygin2016} and can add to the knowledge of general statistical trends of multi-planetary systems (e.g., \citealt{2012PNAS..109.7982S}; \citealt{2016ApJ...825...98H}).  Moreover, high-precision multi-band transit photometry permits exploration of the atmospheric properties of close-in planets, notably conditions related to the presence and potentially the compositions of clouds and hazes (e.g., \citealt{2016Natur.529...59S}).
Hence, we initialized the Transiting Exoplanet Monitoring Project (TEMP, \citealt{2017AJ....154...49W}) to study dozens exoplanet systems which have a lack of follow-up observations and/or show interesting TTV signals. We refine their system parameters, and orbital ephemerides, and characterize their dynamical histories by collecting and analyzing high-precision photometric light curves.

The transiting hot Jupiter HAT-P-25b was discovered by \citet{2012ApJ...745...80Q} under the auspices of the HATNet project. The system comprises a G5 dwarf star and a hot Jupiter, which has a transit period of $P=3.652836\pm0.000019$ days. The host star ($V$=13.19) has an effective temperature of T=5500$\pm$80 K and a mass of 1.010$\pm$0.032 M$_{\odot}$. The mass and radius of the planet were found to be 0.567$\pm$0.022 $\mathrm{M_{J}}$ and $1.190_{-0.056}^{+0.081}$ $\mathrm{R_{J}}$. As one of the first targets in TEMP, the photometric characterization of HAT-P-25 relies primarily on two transit measurements including one incomplete light curve (the two light curves are refitted in this work -- see \S 3) in the discovery paper \citep{2012ApJ...745...80Q}. Furthermore, the mid-transit times of the light curves listed in the Exoplanet Transit Database\footnote{http://var2.astro.cz/ETD/index.php} (ETD) show substantial deviations from the linear ephemeris provided by \citet{2012ApJ...745...80Q}.
As a consequence, follow-up observations are needed to consolidate the system parameters and to improve the overall characterization.

In this work, we present the first transit photometry of HAT-P-25b since the discovery paper, covering eight transits. These new transits, when combined with the published data from \citet{2012ApJ...745...80Q}, allow us to refine the physical and orbital parameters. Based on the analysis of mid-transit times derived from all available follow-up light curves (eight from this work, and two from \citealt{2012ApJ...745...80Q}), we determine an updated linear ephemeris, as well as upper mass limits on potential nearby and long-term perturbers.

We proceed as follows. In {\S} 2, we detail the observations and data reduction. An analysis of the resulting light curves is presented in {\S} 3 and {\S} 4 describes a dynamical analysis of this system, and places the assessment of HAT-P-25 into the broader context provided by the galactic planetary census, before segueing into the final section which contains a brief summary and overview.

\setlength{\tabcolsep}{1.1pt}
\begin{deluxetable*}{cccccccccc}
\tabletypesize{\scriptsize}
\tablewidth{0pt}
\tablecaption{Log of Observations}
\tablehead{
\colhead{Date}  &  \colhead{Time}   &  \colhead{Telescope} & \colhead{Filter} &  \colhead{Number of exposures}  &  \colhead{Exposure time}  &  \colhead{Airmass}  &  \colhead{Moon Phase} &  \colhead{Distance\tablenotemark{a}}&  \colhead{Scatter}\tablenotemark{a}\\
\colhead{(UTC)}  &  \colhead{(UTC)}   &  \colhead{}    & \colhead{}       &  \colhead{}    &  \colhead{(second)}       &  \colhead{}         &  \colhead{} &  \colhead{degree}   &  \colhead{}  }
\startdata
2013 Sep 25                  & { }{ }16:40:25-20:53:39 & { }{ }Weihai $1\,{\rm m}$                  & $V$ &     221 &  50 & 1.02-1.16& 0.00 & 26.00  &  0.0035 \\
2013 Oct 17                  & { }{ }14:58:49-18:32:24 & { }{ }Weihai $1\,{\rm m}$                  & $V$ &     202 &  40 & 1.02-1.18& 0.98 & 43.80  &  0.0053 \\
2013 Nov 19\tablenotemark{b} & { }{ }12:15:18-16:21:57 & { }{ }Xinglong $60/90\,{\rm cm}$ Schmidt   & $R$ &     95  &  120& 1.06-1.37& 0.96 & 27.71  &  0.0030 \\
2013 Nov 30                  & { }{ }10:31:55-14:40:40 & { }{ }Weihai $1\,{\rm m}$                  & $V$ &     229 &  50 & 1.02-1.59& 0.09 & 160.20 &  0.0029 \\
2015 Feb 04                  & { }{ }10:56:25-15:02:29 & { }{ }Xinglong $60/90\,{\rm cm}$ Schmidt   & $R$ &     136 &  80 & 1.04-1.92& 0.99 & 170.00 &  0.0034 \\
2016 Jan 17                  & { }{ }11:07:08-15:29:06 & { }{ }Xinglong $60\,{\rm cm}$              & $R$ &     121 &  100& 1.04-1.55& 0.59 & 20.20  &  0.0027 \\
2016 Jan 28\tablenotemark{b} & { }{ }11:47:06-15:02:28 & { }{ }Xinglong $60/90\,{\rm cm}$ Schmidt   & $R$ &     65  &  150& 1.06-1.67& 0.80 & 125.67 &  0.0039 \\
2016 Nov 04                  & { }{ }16:44:43-21:12:31 & { }{ }Xinglong $60/90\,{\rm cm}$ Schmidt   & $R$ &     84  &  180& 1.04-2.00& 0.25 & 133.33 &  0.0020 \\
\enddata
\tablenotetext{a}{Distance is the mean of the distance between the target and the moon on the sky during the observation. Scatter represents the RMS of the residuals from the best-fitting transit model. }
\tablenotetext{b}{Due to bad weather, the observations were interrupted, resulting in two partial transit light curves. }
\label{obslog}
\end{deluxetable*}
%\clearpage
\section{Observations and Data Reduction}
  Between September 2013 and November 2016, we observed eight transits of HAT-P-25b with three telescopes at two different sites. Three of transits were observed with the 1-m telescope at the Weihai Observatory (WHOT; $122^{\circ} 02^{\prime} 58.6$\arcsec$\mathrm{E}$, $37^{\circ} 32^{\prime} 09.3$\arcsec$\mathrm{N}$) of the Shandong University in China, and others were observed with the 60 cm telescope and the 60/90 cm Schmidt telescope at Xinglong Station ($117^{\circ} 34^{\prime} 30$\arcsec$\mathrm{E}$, $40^{\circ} 23^{\prime} 39$\arcsec$\mathrm{N}$) of the National Astronomical Observatories of China (NAOC).
\subsection{Weihai Observatory}

Three Johnson $V$ band transits of HAT-P-25b were obtained at Weihai Observatory in 2013, on September 25, October 17 and November 30. With a 2K$\times$2K imaging array (13.5$\times$13.5 $\mu$m pixel$^{-1}$), the telescope provides a field of view of $12^{\prime}\times12^{\prime}$, and a pixel scale of 0.35$\arcsec$  pixel$^{-1}$. The full technical details of this telescope are given \citet{2014RAA....14..719H}.

 In each run, exposure times were held fixed to avoid adverse systematic effects on the measurements of mid-transit times. The time on the telescope and CCD control computers were GPS-synchronized at a one-minute cadence. The HJD time stamps in each FITS header log the mid-exposure time, and were logged from the synchronized system time following the UTC time standard. For the accurate timing studies presented later in this work, we converted the HJD time stamps in the UTC time standard to BJD time stamps valid in the TDB time standard. The precise time management techniques that we used are described in \citet{2010PASP..122..935E}.

Using standard procedures, all data were debiased and flat-fielded. Aperture photometry was then obtained using the Source Extractor Software Package \citep{1996A&AS..117..393B}. Transit light curves were obtained using differential photometry. The choice of apertures and photometric comparison stars were adjusted manually in order to produce the lowest scatter among observations taken out of transit.

\subsection{Xinglong Station, National Astronomical Observatory}
Between 2013 November and 2016 November, four Johnson $R$ band transits of HAT-P-25b were obtained with the 60/90 cm Schmidt telescope at Xinglong Station of the NAOC of China, and an additional transit was monitored with the R-band using the 60 cm telescope. The 60/90 cm Schmidt telescope is equipped with a 4K$\times$4K CCD, which gives a $94^{\prime}\times94^{\prime}$ field of view with a pixel scale of 1.38$\arcsec$ pixel$^{-1}$. The 60 cm telescope is equipped with a 1K$\times$1K CCD, which provides a $17^{\prime}\times17^{\prime}$ effective field of view with a pixel scale of 1.00$\arcsec$ pixel$^{-1}$. More instrumental details for these telescopes can be found in \citet{1999PASP..111..909Z} and \citet{2009PASJ...61.1211Y}.

Due to bad weather, the observations on Nov 19, 2013, and Jan 28, 2016, were interrupted, which resulted in two partial transit light curves.

 We used the same strategy described in \S 2.1 to handle the data from the observations performed at Xinglong station. A detailed summary of observations is presented in Table~\ref{obslog}. The resulting transit curves are listed in Table~\ref{resultinglightcurves.}, and are shown in Figures~\ref{TTVglobal} and~\ref{lightcurvesFIG}.

\setlength{\tabcolsep}{1.1pt}
\begin{deluxetable}{lcccc}
\tabletypesize{\scriptsize}
\tablewidth{0pt}
\tablecaption{Photometry of HAT-P-25}
\tablehead{\colhead{$ \mathrm{BJD}_\mathrm{TDB}\tablenotemark{a}$} & \colhead{Relative Flux} & \colhead{Uncertainty}& \colhead{Filter}}
\startdata
2456561.199442 &  1.0018    & 0.0028 &    R    \\
2456561.200288 &  1.0018    & 0.0028 &    R    \\
2456561.201028 &  0.9954    & 0.0028 &    R   \\
2456561.201881 &  0.9932    & 0.0028 &    R   \\
2456561.202627 &  0.9954    & 0.0028 &    R   \\
2456561.203472 &  1.0008    & 0.0028 &    R   \\
2456561.204225 &  1.0008    & 0.0028 &    R   \\
2456561.205068 &  0.9954    & 0.0028 &    R   \\
2456561.205809 &  1.0014    & 0.0028 &    R   \\
... &  ...    & ... &    ...\\
\enddata
\tablenotetext{a}{Time stamps throughout the paper have been converted to $ \mathrm{BJD}_\mathrm{TDB}$.  }
\label{resultinglightcurves.}
\end{deluxetable}

%\begin{verbatim}

%\end{verbatim}
%\begin{figure}
%\begin{center}
%\includegraphics[width=0.5\textwidth]{TTV.eps}
%\caption{EXOFAST}
%\end{center}
%\end{figure}
%\begin{figure}
%\begin{center}
%\includegraphics[width=0.5\textwidth]{RV.eps}
%\caption{EXOFAST}
%\end{center}
%\end{figure}

\section{Light curve analysis}
\label{Lightcurevanalysis}

To re-estimate the global parameters, we employed Multi-EXOFAST\footnote{A description of the procedure can be found at \url{http://astroutils.astronomy.ohio-state.edu/exofast/.}} \citep{2013PASP..125...83E}, a speed-optimized suite of exoplanet model-fitting software written in IDL which can fit multiple follow-up transit light curves in different filter bands and multiple RV telescope data sets \citep{ collins2017}. The package is able to fit transit data and RVs simultaneously, which can improve the quality of both fit types and provide a clearer picture of the system under consideration \citep{2013PASP..125...83E}. In our treatment, each data set was first fitted separately to scale the uncertainties and to derive a preliminary best fit. Then, based on the Differential Evolution Markov Chain Monte Carlo (DE-MCMC; \citealt{DBLP:journals/sac/Braak06}), the package was employed to carry out a global fit to all data sets. This permitted refinement of the best joint fit. Stellar parameters were calculated with the help of Torres relations \citep{2008ApJ...677.1324T}. By using standard MCMC techniques to evaluate the posterior density, we determined robust uncertainty estimates for the parameters. As described in  \citet{2013PASP..125...83E}, a given Markov chain is considered to be converged when both the number of independent draws is greater than 1000 and the Gelman-Rubin statistic is less than 1.01 for all parameters. During fitting, Markov chains that satisfy these criteria six consecutive times are considered to be well-mixed.

\begin{figure}[hpb]

\includegraphics[width=0.5\textwidth]{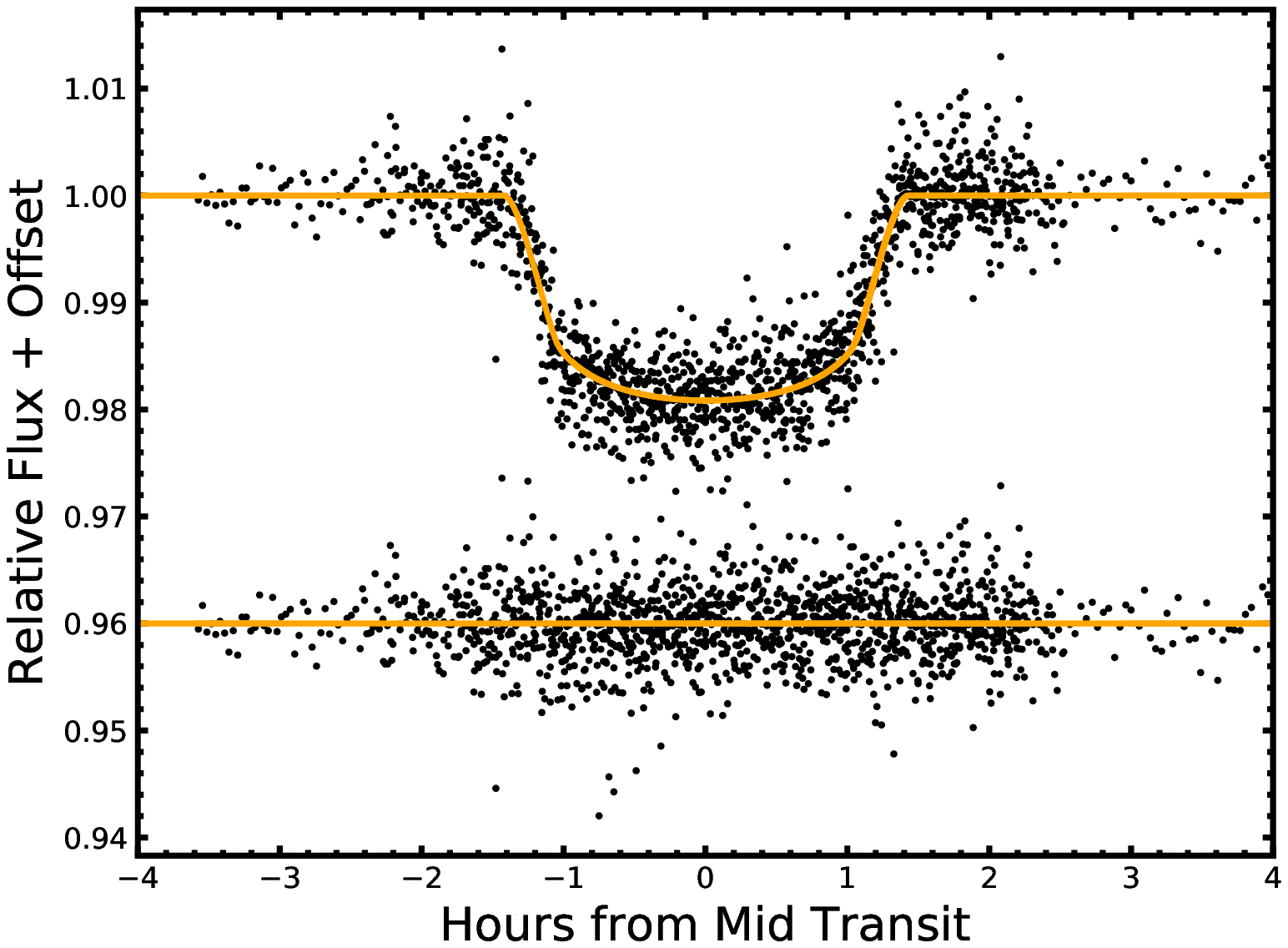}
\begin{center}
\caption{Combined photometry of HAT-P-25, shown as a function of orbital phase. To estimate the system parameters, we fit the combined light curves along with the radial velocity measurements (not shown). The best-fitting model to all of the data is plotted as solid orange line and the residuals (once the model has been removed) are shown in the lower part of the figure.}
\label{TTVglobal}
\end{center}
%\end{figure}

%\begin{figure}[h]
\begin{center}
%\flushright
\includegraphics[width=0.525\textwidth]{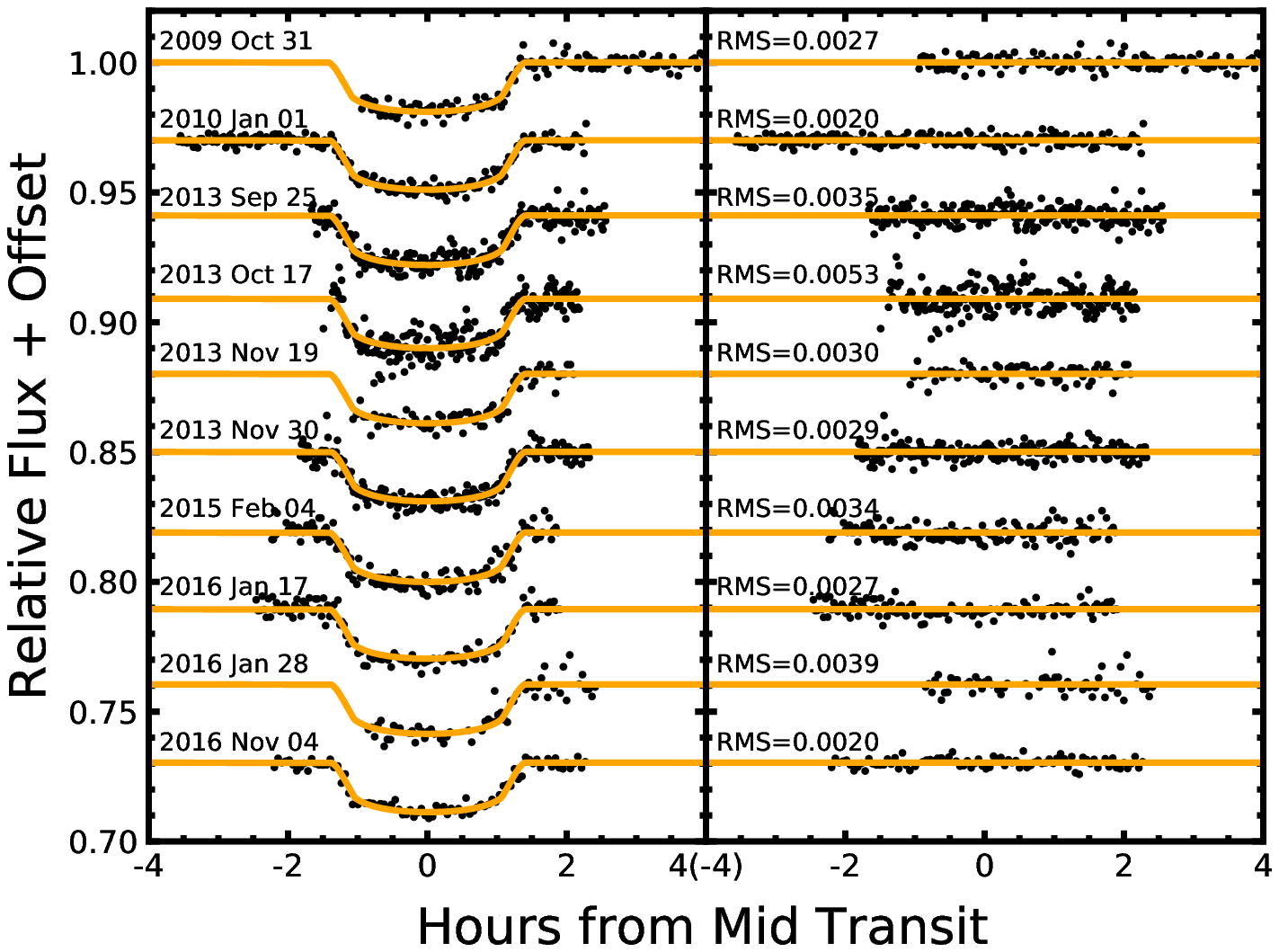}
\caption{Light curves of the HAT-P-25b transit obtained in individual runs. The Top two light curves are from \citet{2012ApJ...745...80Q}, and the bottom eight light curves belong to this work. Based on JKTEBOP code, the mid-transit times are estimated by analyzing these data. The best-fitting models, from the joint analysis of the RV data and the ensemble photometric light curves, are plotted as solid orange lines. The residuals of fits and their RMS errors from JKTEBOP are displayed on the right panel, in the same order as the light curves. For clarity, we make offset of light curves and residuals artificial. More details of each light curves can be seen in Table~\ref{obslog}.  }
\label{lightcurvesFIG}
\end{center}
\end{figure}

For comparison purposes, we performed global fitting for three different incarnations of the joint data set. In the first incarnation, we used the eight new light curves obtained from this work in conjunction with the RVs from discovery work (a set we call ``8 new light curves + RVs''). In the second incarnation we used the two extant light curves from the literature in conjunction with RVs from the discovery work (a set we call ``2 literature light curves + RVs''). In the third incarnation we used Multi-EXOFAST to deal with all ten above-mentioned light curves in three different filter bands and the RVs from discovery work (a set we call ``All light curves + RVs''). The parameters we used in order to initialize the fits are from \citet{2012ApJ...745...80Q}. The results from the three different global fittings and their errors are presented in Table~\ref{global}. After making a comparison between the fits (details can be seen in \S 4.1), we selected the parameters from the ``All light curves + RVs'' set as the best representation of the global parameters. The fitting results are shown in Figure~\ref{lightcurvesFIG}.

%Overall, the mean of RMSs of 8 new transit light curves (3.225 mmag) is bigger than 2 literature transit light curves (2.355 mmag), which means the quality of new transit light curves is little worse. The RMSs are presented in Table~\ref{obslog} and Figure~\ref{lightcurvesFIG}.

With the best global parameters derived from the ``All light curves + RVs'' data set in hand, we applied the JKTEBOP routine\footnote{Code is available in its entirety at http://www.astro.keele.ac.uk/jkt/codes/jktebop.html.} (\citealt{2004MNRAS.349..547S}; \citealt{2004MNRAS.351.1277S}), a fast procedure that can analyze light curves of detached eclipsing binaries and transiting extra-solar planetary systems. This allowed us to accurately measure mid-transit times of the eight new light curves from this work and the two acquired from the discovery paper \citep{2012ApJ...745...80Q}. In the JKTEBOP fitting, we fixed all the parameters derived from previously global fitting and just considered the mid-transit times and baseline fluxes as free parameters for each light curve to float. Using Levenberg-Marquardt optimization, we derived mid-transit times. Furthermore, through comparing uncertainties generated with the residual-permutation algorithm and with Monte Carlo simulations (10,000 trials), we consistently chose the larger of the obtained uncertainties of mid-transit times to secure a conservative estimate. The mid-transit times derived using JKTEBOP are listed in Table~\ref{midtimes}.

\section{Result and Analysis}

\subsection{System Parameters}
\begin{figure}[t!]
\begin{center}
\includegraphics[width=0.50\textwidth]{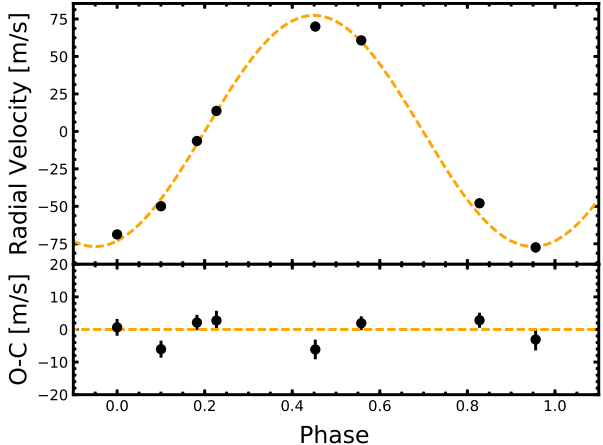}
\caption{Keck/HIRES RV measurements of HAT-P-25 from \citet{2012ApJ...745...80Q}, along with the best-fit model from the joint modeling of RVs and light curves. The best-fit is plotted as dashed orange line. The O-C residuals are displayed on the bottom panel, which has an RMS scatter $\sigma= 3.89 {ms}^{-1}$.}
\label{RVFIG}
\end{center}
\end{figure}

Based on the analysis described in \S 3, the updated physical and orbital parameters for the HAT-P-25 system, along with their errors, are presented in Tables~\ref{global}. The tables also include estimates stemming from the previous work \citep{2012ApJ...745...80Q}. The best-fitting model of transit and RV are shown in Figures~\ref{TTVglobal} and~\ref{RVFIG}.

The results of the three different global fits are mutually consistent. Moreover, compared with the quality of the parameters in the other two fittings, the ``All light curves + RVs'' fit indicates that the accuracy of 28 of the 46 parameters has improved; all important parameters were improved except stellar mass, effective temperature, metallicity, argument of periastron, planetary mass, ingress/egress duration (details can be seen in Table~\ref{global}). Therefore, the result of the global fitting, the ``All light curves + RVs'' model, is regarded as the best global fitting result. Extrapolating from the RMS residuals of individual fits shown in Figure~\ref{lightcurvesFIG} and the observation logs, we can conclude that with more favorable sky conditions and with the use longer exposure times (in conjunction with appropriate defocusing) one might be able to obtain lower RMS\footnote{Because the light curves of `2013 Nov 19' and `2016 Jan 28' are partial, we performed a Pearson correlation analysis the corpus of data exclusive of these light curves. The Pearson correlation coefficient between exposure times and RMS is $r=-0.74$, whereas that between the moon-target distance and the RMS is $r=-0.36$.}. We have also carried out a similar model calculation that retains only the $V/Sloan i/R$-filter data. For that case, the planetary and stellar radius ratios $R_{P}/R_{*}$ in each band were found to be within the radius measurement presented in Table~\ref{bandvsradii}. From this we conclude that the current data do not allow for differentiation between pass-band dependent radius measurements. In summary, our work provides the following updates to the characterization of the system provided in the discovery paper:

Given that the same RVs were used, it comes as no surprise that we find all Doppler velocimetric properties consistent with those of discovery work \citep{2012ApJ...745...80Q}. The RV parameters derived from our work are all in agreement with those from discovery work to within 0.61$\sigma$.

The transit parameters that we derive are all consistent with the discovery work, and all of the parameters other than the inclination have had their uncertainties reduced. Given that the identical stellar spectroscopic parameters were used here and in the discovery work, there is full consistency with \citet{2012ApJ...745...80Q}.

Perhaps the most important update from our analysis is the significant improvement to the period, which stems from the use of a more extensive collection of mid-transit times (see the description in \S 4.2).

Based on the above, the unsurprising conclusion is that the physical and orbital parameters for HAT-P-25b agree with those of discovery work within the uncertainties.

\subsection{Transit Timing}
\begin{figure}
\begin{center}
\includegraphics[width=0.5\textwidth]{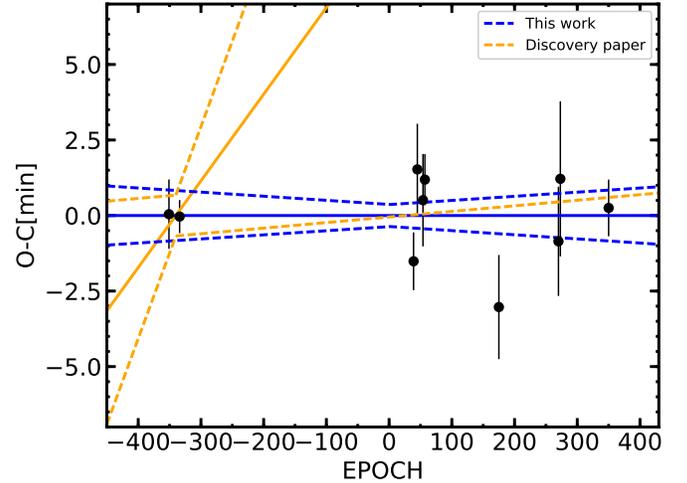}
\caption{ Observed minus calculated mid-transit times for HAT-P-25b. Adopting the updated linear ephemeris, transit timing residuals calculated with all transit epochs are shown. The first two mid-transit times are from \citet{2012ApJ...745...80Q}, and others belong to this work. The solid blue line represents zero deviation from the predicted time of transit. The blue dashed line represents the propagation of $\pm$1$\sigma$ errors associated with the calculated orbital period. Based on the linear ephemeris calculated using the two transit light curves from \citet{2012ApJ...745...80Q}, a solid yellow line and dash yellow lines are plotted. These show the effect of using the imperfectly determined period. }
%The slope of blue dash line results for a mis-determined period.
\label{TTVFIG}
\end{center}
\end{figure}

Mid-transit times, T$_{C}$, derived from each new and literature light curves, are listed in Table~\ref{midtimes}. To calculate an updated linear ephemeris, that can be employed to predict the transit time accurately and to analyze TTVs, we employed a weighted least square method to fit all ten mid-transit times (T$_{C}$) with a linear function of transit epoch number (E),
\begin{equation}
    T_{C[E]}=T_{C[0]}+E \times P,
\end{equation}

where $T_{C[E]}$ is the mid-transit time, $T_{C[0]}$ is mid-transit time of the reference epoch and $P$ is period. The results are $T_{C[0]}$=2456418.80996$\pm$0.00025 [$\mathrm{BJD}_\mathrm{TDB}$] and $P$=3.65281572$\pm$0.00000095 days. This fit has a reduced $\chi^{2}$=1.60 (RMS=80 s). The new orbital period of HAT-P-25b is twenty times more precise than the period reported in discovery work, an improvement enabled by our long-term series of follow-up observations. We chose the middle epoch within our observations as the reference epoch to minimize covariance with the orbital period. Furthermore, the updated linear ephemeris not only fits the ten mid-transit times in this work well ($\chi_{\rm reduced}^{2}$=1.60), but also fits the data of ETD well, giving $\chi_{\rm reduced}^{2}=9.76$ whereas the linear ephemeris of discovery work gives $\chi_{\rm reduced}^{2}=130.45$. Figure~\ref{TTVFIG} displays the deviation of mid-transit times for HAT-P-25b from our updated linear ephemeris, showing no significant timing variations at a level above 80s.

Given the lack of significant timing variations, we can calculate the upper mass-limits of potential nearby perturbers (\S  4.3) and long-term perturbers (\S  4.4). The determinations are based on the analysis of TTVs caused by planet-perturbr interaction ($\mathrm{TTV_{G}}$) and the perturber-induced stellar barycentric movement (known as light travel time effect, $\mathrm{TTV_{LTTE}}$) respectively.

\setlength{\tabcolsep}{1.1pt}
\begin{longrotatetable}
\begin{deluxetable*}{lccccccc}
\tabletypesize{\scriptsize}
\tablewidth{0pt}
\tablecaption{System parameters for HAT-P-25\label{global}}
\tablehead{\colhead{~~~Parameter} & \colhead{Units} & \colhead{Priors\tablenotemark{a}}& \colhead{8 new LCs + RVs}& \colhead{2 literature LCs + RVs}& \colhead{All light curves + RVs}& \colhead{\citet{2012ApJ...745...80Q}\tablenotemark{b}}& \colhead{agreement($\sigma$)\tablenotemark{c}}}
\startdata
\sidehead{Stellar Parameters:}
                           ~~~$M_{*}$\dotfill &Mass (\msun)\dotfill & $1.010\pm0.032$                     & $1.018_{-0.053}^{+0.055}$             & $1.007_{-0.052}^{+0.054}$                & $1.012_{-0.051}^{+0.053}$                  & $1.010\pm0.032$                                  &0.03\\
                         ~~~$R_{*}$\dotfill &Radius (\rsun)\dotfill & $0.947_{-0.041}^{+0.044}$           & $0.930_{-0.038}^{+0.039}$             & $0.926_{-0.041}^{+0.043}$                & $0.919\pm0.034$                            & $0.959_{-0.037}^{+0.054}$                        &0.80\\
                     ~~~$L_{*}$\dotfill &Luminosity (\lsun)\dotfill & ...                                 & $0.731_{-0.076}^{+0.086}$             & $0.692_{-0.077}^{+0.088}$                & $0.705_{-0.070}^{+0.076}$                  & $0.75\pm0.10$                                    &0.36\\
                         ~~~$\rho_*$\dotfill &Density (cgs)\dotfill & ...                                 & $1.79_{-0.17}^{+0.19}$                & $1.79_{-0.19}^{+0.22}$                   & $1.84_{-0.15}^{+0.17}$                     &...                                               &...\\
              ~~~$\log(g_*)$\dotfill &Surface gravity (cgs)\dotfill & $4.48\pm0.04$                       & $4.509\pm0.028$                       & $4.508\pm0.032$                          & $4.516_{-0.025}^{+0.026}$                  & $4.48\pm0.04$                                    &0.76\\
              ~~~$\teff$\dotfill &Effective temperature (K)\dotfill & $5500\pm80$                         & $5540_{-78}^{+79}$                    & $5476\pm80$                              & $5519_{-76}^{+78}$                         & $5500\pm80$                                      &0.17\\
                              ~~~$\feh$\dotfill &Metalicity\dotfill & $0.31\pm0.08$                       & $0.286_{-0.080}^{+0.079}$             & $0.314_{-0.078}^{+0.080}$                & $0.294\pm0.080$                            & $0.31\pm0.08$                                    &0.14\\
\sidehead{Planetary Parameters:}
                               ~~~$e$\dotfill &Eccentricity\dotfill & $0.032\pm0.022$                     & $0.024_{-0.015}^{+0.022}$              & $0.025_{-0.016}^{+0.022}$                &  $0.023_{-0.014}^{+0.022}$                 & $0.032\pm0.022$                                  &0.29\\
    ~~~$\omega_*$\dotfill &Argument of periastron (degrees)\dotfill & $271\pm117$                         & $-74_{-16}^{+50}$                      & $286_{-15}^{+46}$                        &  $287_{-17}^{+52}$                         & $271\pm117$                                      &0.14\\
                              ~~~$P$\dotfill &Period (days)\dotfill & $3.652836\pm0.000019$               & $3.6528140_{-0.0000018}^{+0.0000017}$  & $3.6528153_{-0.0000079}^{+0.0000080}$    &  $3.65281514_{-0.00000075}^{+0.00000076}$  & $3.652836\pm0.000019$                            & 1.10\\
                       ~~~$a$\dotfill &Semi-major axis (AU)\dotfill & $0.04658\pm0.000776$                & $0.04669\pm0.00082$                    & $0.04653\pm0.00082$                      &  $0.04660_{-0.00080}^{+0.00081}$           & $0.0466\pm0.0005$                                &0.00\\
                             ~~~$M_{P}$\dotfill &Mass (\mj)\dotfill & ...                                 & $0.571_{-0.022}^{+0.023}$              & $0.568\pm0.022$                          &  $0.569_{-0.022}^{+0.023}$                 & $0.567\pm0.022$                                  &0.06\\
                           ~~~$R_{P}$\dotfill &Radius (\rj)\dotfill & $1.190_{-0.056}^{+0.081}$           & $1.154_{-0.056}^{+0.057}$              & $1.143_{-0.060}^{+0.063}$                &  $1.135\pm0.048$                           & $1.190_{-0.056}^{+0.081}$                        &0.75\\
                       ~~~$\rho_{P}$\dotfill &Density (cgs)\dotfill & $0.42\pm0.070$                      & $0.462_{-0.057}^{+0.067}$              & $0.471_{-0.064}^{+0.075}$                &  $0.483_{-0.051}^{+0.059}$                 & $0.42\pm0.070$                                   &0.73\\
                  ~~~$\log(g_{P})$\dotfill &Surface gravity\dotfill & ...                                 & $3.027_{-0.037}^{+0.038}$              & $3.032\pm0.042$                          &  $3.039_{-0.032}^{+0.033}$                 & $3.0_{-0.06}^{+0.04}$                            &0.76\\
           ~~~$T_{eq}$\dotfill &Equilibrium Temperature (K)\dotfill & ...                                 & $1191_{-26}^{+28}$                     & $1177_{-29}^{+30}$                       &  $1182\pm25$                               & $1202\pm36$                                      &0.46\\
                       ~~~$\Theta$\dotfill &Safronov Number\dotfill & ...                                 & $0.0454_{-0.0023}^{+0.0025}$           & $0.0458_{-0.0025}^{+0.0027}$             &  $0.0461_{-0.0021}^{+0.0023}$              & $0.044\pm0.003$                                  &0.57\\
               ~~~$\fave$\dotfill &Incident flux (\fluxcgs)\dotfill & ...                                 & $0.457_{-0.039}^{+0.044}$              & $0.436_{-0.041}^{+0.046}$                &  $0.442_{-0.036}^{+0.039}$                 & $0.472\pm0.58$                                   &0.05\\
\sidehead{RV Parameters:}
                              ~~~$e\cos\omega_*$\dotfill & \dotfill & ...                                 & $0.0060_{-0.0061}^{+0.0069}$          & $0.0064_{-0.0062}^{+0.0068}$             & $0.0062_{-0.0062}^{+0.0068}$               & $0.008\pm0.012$                                   &0.13\\
                              ~~~$e\sin\omega_*$\dotfill & \dotfill & ...                                 & $-0.022_{-0.024}^{+0.020}$            & $-0.023_{-0.024}^{+0.020}$               & $-0.020_{-0.024}^{+0.019}$                 & $-0.020\pm0.034$                                  &0.00\\
           ~~~$T_{P}$\dotfill &Time of periastron (\bjdtdb)\dotfill & ...                                 & $2455200.77_{-0.17}^{+0.51}$          & $5200.77_{-0.16}^{+0.47}$                & $5200.78_{-0.18}^{+0.53}$                  & ...                                               &...\\
                    ~~~$K$\dotfill &RV semi-amplitude (m/s)\dotfill & $74.3\pm2.4$                        & $74.5\pm1.4$                          & $74.5\pm1.4$                             & $74.5\pm1.4$                               & $74.3\pm2.4$                                      &0.14\\
                 ~~~$M_P\sin i$\dotfill &Minimum mass (\mj)\dotfill & $0.56683\pm0.026307$                & $0.571_{-0.022}^{+0.023}$             & $0.567\pm0.022$                          & $0.569_{-0.022}^{+0.023}$                  & ...                                               &...\\
                       ~~~$M_{P}/M_{*}$\dotfill &Mass ratio\dotfill & ...                                 & $0.000536\pm0.000014$                 & $0.000538\pm0.000014$                    & $0.000537\pm0.000014$                      & ...                                               &...\\
               ~~~$\gamma$\dotfill &Systemic velocity (m/s)\dotfill & ...                                 & $1.1\pm1.0$                           & $1.1\pm1.0$                              & $1.0\pm1.0$                                & ...                                               &...\\
              ~~~$\dot{\gamma}$\dotfill &RV slope (m/s/day)\dotfill & ...                                 & $-0.046_{-0.046}^{+0.045}$            & $-0.047\pm0.046$                         & $-0.047_{-0.046}^{+0.045}$                 & ...                                               &...\\
\sidehead{Primary Transit Parameters:}
~~~$R_{P}/R_{*}$\dotfill &Radius of planet in stellar radii\dotfill & ...                                 & $0.1275_{-0.0014}^{+0.0013}$          & $0.1269\pm0.0015$                        & $0.1269\pm0.0011$                          & $0.1275\pm0.0024$                                 &0.23\\
     ~~~$a/R_{*}$\dotfill &Semi-major axis in stellar radii\dotfill & $10.46_{-0.55}^{+0.38}$             & $10.80_{-0.35}^{+0.37}$               & $10.81_{-0.40}^{+0.42}$                  & $10.90_{-0.31}^{+0.33}$                    & $10.46_{-0.55}^{+0.38}$                           &0.9\\
              ~~~$u_1$\dotfill &linear limb-darkening coeff\dotfill & ...                                 & ...\tablenotemark{d}                  & $0.373\pm0.014$                          & ...\tablenotemark{e}                       & $0.3287$                                          &...\\
           ~~~$u_2$\dotfill &quadratic limb-darkening coeff\dotfill & ...                                 & ...\tablenotemark{d}                  & $0.2473_{-0.0084}^{+0.0077}$             & ...\tablenotemark{e}                       & $0.3039$                                          &...\\
                      ~~~$i$\dotfill &Inclination (degrees)\dotfill & $87.6\pm0.5$                        & $88.09_{-0.40}^{+0.53}$               & $88.00_{-0.48}^{+0.62}$                  & $88.22_{-0.36}^{+0.45}$                    & $87.6\pm0.5$                                      &1.01\\
                           ~~~$b$\dotfill &Impact Parameter\dotfill & $0.45607_{-0.066586}^{+0.068411}$   & $0.369_{-0.096}^{+0.067}$             & $0.386_{-0.11}^{+0.079}$                 & $0.347_{-0.082}^{+0.061}$                  & $0.456_{-0.098}^{+0.073}$                          &0.94\\
                         ~~~$\delta$\dotfill &Transit depth\dotfill & $0.016526\pm0.000612$               & $0.01626\pm0.00035$                   & $0.01610_{-0.00037}^{+0.00038}$          & $0.01610\pm0.00027$                        &...                                                &...\\
                ~~~$T_{FWHM}$\dotfill &FWHM duration (days)\dotfill & ...                                 & $0.10258\pm0.00055$                   & $0.10178_{-0.00086}^{+0.00085}$          & $0.10235_{-0.00047}^{+0.00048}$            &...                                                &...\\
          ~~~$\tau$\dotfill &Ingress/egress duration (days)\dotfill & ...                                 & $0.0152\pm0.0012$                     & $0.0153_{-0.0013}^{+0.0015}$             & $0.01485_{-0.00091}^{+0.00094}$            & $0.0163\pm0.0018$                                 &0.71 \\
                 ~~~$T_{14}$\dotfill &Total duration (days)\dotfill & $0.1174\pm0.0017$                   & $0.1178\pm0.0012$                     & $0.1170_{-0.0013}^{+0.0014}$             & $0.11721_{-0.00091}^{+0.00093}$            & $0.1174\pm0.0017$                                 &0.1\\
      ~~~$P_{T}$\dotfill &A priori non-grazing transit prob\dotfill & ...                                 & $0.0791_{-0.0036}^{+0.0035}$          & $0.0790\pm0.0037$                        & $0.0785_{-0.0035}^{+0.0032}$               &...                                                &...\\
                ~~~$P_{T,G}$\dotfill &A priori transit prob\dotfill & ...                                 & $0.1021_{-0.0047}^{+0.0046}$          & $0.1019\pm0.0048$                        & $0.1013_{-0.0045}^{+0.0042}$               &...                                                &...\\
                            ~~~$F_0$\dotfill &Baseline flux\dotfill & ...                                 & $1.00018\pm0.00016$                   & $1.00001\pm0.00016$                      & $1.00009\pm0.00011$                        &...                                                &...\\
\sidehead{Secondary Eclipse Parameters:}
              ~~~${T_S}\tablenotemark{f}$\dotfill &Time of eclipse (\bjdtdb)\dotfill & ...                & $5200.611_{-0.014}^{+0.016}$          & $5200.611_{-0.014}^{+0.016}$             & $5200.611_{-0.014}^{+0.016}$              & $5178.698\pm0.027$                                &...\\
                       ~~~$b_{S}$\dotfill &Impact parameter\dotfill & ...                                 & $0.351_{-0.088}^{+0.062}$             & $0.366_{-0.10}^{+0.071}$                 & $0.331_{-0.076}^{+0.057}$                  &...                                                &...\\
              ~~~$T_{S,FWHM}$\dotfill &FWHM duration (days)\dotfill & ...                                 & $0.0989_{-0.0039}^{+0.0034}$          & $0.0980_{-0.0037}^{+0.0034}$             & $0.0988_{-0.0040}^{+0.0033}$               &...                                                &...\\
        ~~~$\tau_S$\dotfill &Ingress/egress duration (days)\dotfill & ...                                 & $0.0144_{-0.0010}^{+0.0011}$          & $0.0144_{-0.0012}^{+0.0013}$             & $0.01409_{-0.00091}^{+0.00096}$            & $0.0154\pm0.0018$                                 &0.64\\
               ~~~$T_{S,14}$\dotfill &Total duration (days)\dotfill & ...                                 & $0.1134_{-0.0045}^{+0.0038}$          & $0.1125_{-0.0044}^{+0.0038}$             & $0.1130_{-0.0046}^{+0.0037}$               & $0.1137\pm0.0060$                                 &0.10\\
      ~~~$P_{S}$\dotfill &A priori non-grazing eclipse prob\dotfill & ...                                 & $0.0828_{-0.0027}^{+0.0026}$          & $0.0829_{-0.0031}^{+0.0032}$             & $0.0820\pm0.0021$                          &...                                                &...\\
                ~~~$P_{S,G}$\dotfill &A priori eclipse prob\dotfill & ...                                 & $0.1070_{-0.0037}^{+0.0036}$          & $0.1070_{-0.0043}^{+0.0044}$             & $0.1058\pm0.0029$                          &...                                                &...\\
\enddata
\tablenotetext{a}{Priors for the transit fitting are available at http://exoplanets.org/csv-files/exoplanets.csv.}
\tablenotetext{b}{`...' indicates the parameter was not determined in the discovery work.}
\tablenotetext{c}{The agreement between `All LCs + RVs' and the discovery work \citep{2012ApJ...745...80Q} is calculated.}
\tablenotetext{d}{For $R$ band, $u_{1}$=$0.432_{-0.016}^{+0.017}$, $u_{2}$=$0.2461_{-0.0100}^{+0.0095}$; for $V$ band, $u_{1}$=$0.542\pm0.020$, $0.206_{-0.014}^{+0.013}$.}
\tablenotetext{e}{For $R$ band, $u_{1}$=$0.437_{-0.016}^{+0.017}$, $u_{2}$=$0.2433_{-0.010}^{+0.0094}$; for $Sloan i$ band, $u_{1}$=$0.365\pm0.014$, $u_{2}$=$0.2519_{-0.0079}^{+0.0072}$; for $V$ band, $u_{1}$=$0.548\pm0.020$, $u_{2}$=$0.202\pm0.014$.}
\tablenotetext{f}{For clarity, $T_S$ are given in the form of $ \mathrm{BJD}_\mathrm{TDB}$-2450000.}

\end{deluxetable*}

\end{longrotatetable}

%including
%1). how do you get transit timing, new period, and transit timing variations.
%Figure 4 (Time vs transit timing variations)
%Table 4 (transit timing variation data)
%2). do you think there are significant TTV signals or not (usually not).

\setlength{\tabcolsep}{1.1pt}
\begin{deluxetable}{lcccc}
\tabletypesize{\scriptsize}
\tablewidth{0pt}
\tablecaption{Values of $R_{P}/R_{*}$ for each of the fliter data\label{bandvsradii}}
\tablehead{\colhead{ Passband}&\colhead{$R_{P}/R_{*}$}&\colhead{$R_{P}/R_{*}$ of All LCs + RVs}&\colhead{agreement($\sigma$)\tablenotemark{a}}}
\startdata
$V$               &  $0.1270\pm0.0021 $               &        & 0.04         \\
$Sloan i$         &  $0.1269\pm0.0015 $               &  $0.1269_{-0.0016}^{+0.0011}$   & 0.00         \\
$R$               &  $0.1278_{-0.0019}^{+0.0022}$     &        & 0.41         \\
\enddata
\tablenotetext{a}{The agreement between `All LCs + RVs' and each filter data is calculated. }

\end{deluxetable}

\subsection{Orbital stability and mass-limit of a nearby perturber}

Our timing study allows us to place an upper mass-limit on hypothetical perturbing planets located on orbits either interior or exterior to the orbit of the transiting planet. This technique is a potentially promising method to detect additional companions in a known transiting system \citep{agol2005,holman2005,nesmor2008}. Upper mass-limit determinations are accomplished via numerical orbit integrations as has previously ben done by several studies \citep{hoyer2011,hoyer2012,2017AJ....154...49W}. We have modified the Fortran-based MECHANIC \citep{slonina2015} orbital integration package to detect and accurately calculate transit events in the presence of a selected perturbing planet. This is done via a series of iterative back-and-forth integrations once the transiting planet passes the face of the host star. The integrator\footnote{odex.f: http://www.unige.ch/$\sim$hairer/prog/nonstiff/odex.f} is based on an extrapolation-algorithm adopting the explicit midpoint rule. Automatic step-size control and order selection are key features that permit this algorithm to be both robust and accurate. Integration control parameters were set slightly above the machine precision for maximum accuracy (at the expense of computing time). The MECHANIC package utilizes OpenMPI\footnote{http://www.open-mpi.org/} allowing the simultaneous spawning of hundreds of parallel integrations. We took advantage of the availability of computing power and chose a direct brute-force (but highly robust) approach, in calculating root-mean-square statistics for numerical transit timing data. Within the frame-work of the three-body problem we integrated the equations of motion and recorded the transit number and time in the event of a transit by iteration. In these calculations we used both the best-fit radii of the planet and its host star. This generated a series of mid-transit times to which we found an analytic least-squares fit enabling the calculation of the RMS statistic.  We have tested this procedure by reproducing the TTV signal shown by Figure 1 in \citet{nesmor2008}. The difference between TTV signals in \citet{nesmor2008} and in our code test is on a 1-second level or below. This test was also carried out previously and additional information on the calculation of TTVs is given in \citet{hinse2015}. This is done on a grid of masses and semi-major axes for the perturbing planet. At the same time (same grid point) we calculated the MEGNO factor \citep{cinc2000,megno2001,cinc2003,hinse2010} during the integration. MEGNO provides information on the degree of chaos present in the system. A MEGNO close to 2 indicates quasi-periodic (usually regular) behavior while values substantially larger than 2 indicate chaos, which is often accompanied by large-scale orbital instabilities. The MEGNO technique's main advantage is the localization of orbital mean-motion resonances, and we utilize this advantage in the present work.

\begin{figure*}
\centering
{\includegraphics[width=0.49\textwidth]{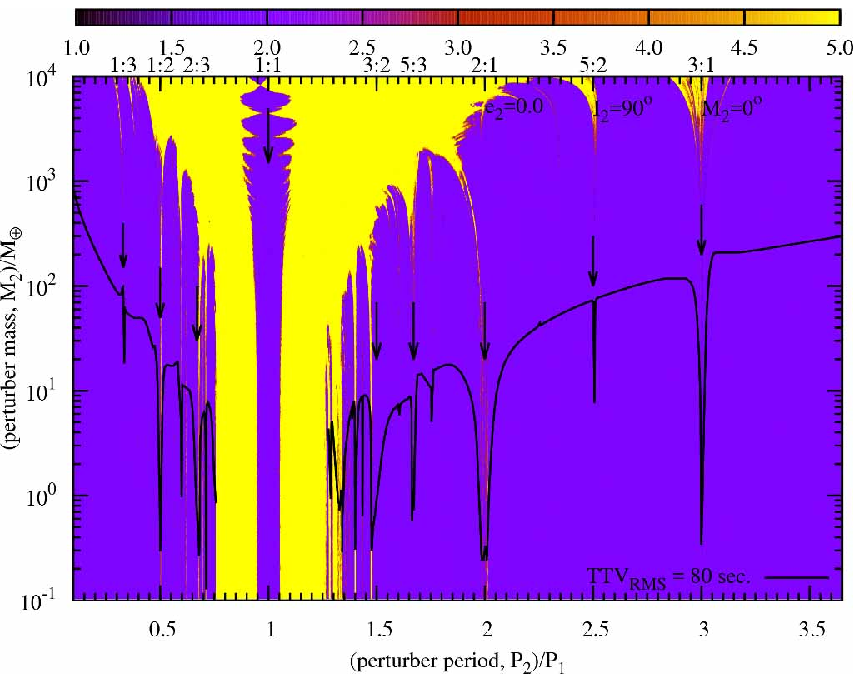}}
{\includegraphics[width=0.49\textwidth]{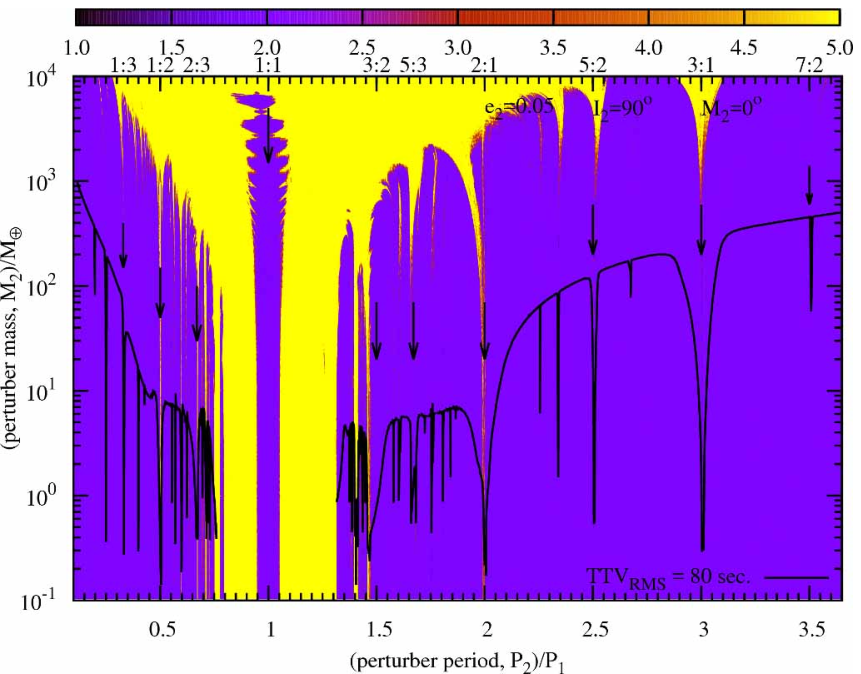}}
{\includegraphics[width=0.49\textwidth]{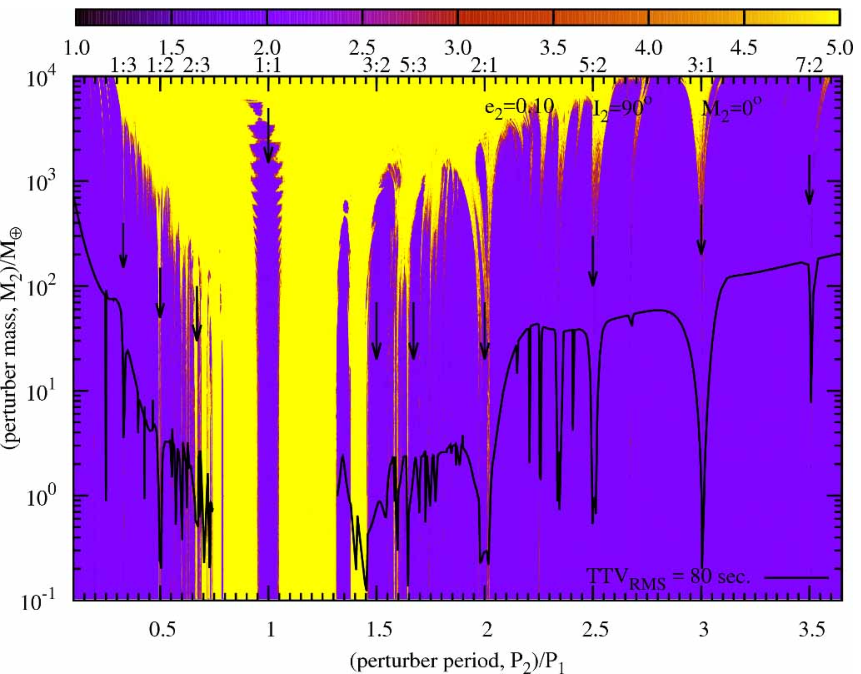}}
{\includegraphics[width=0.49\textwidth]{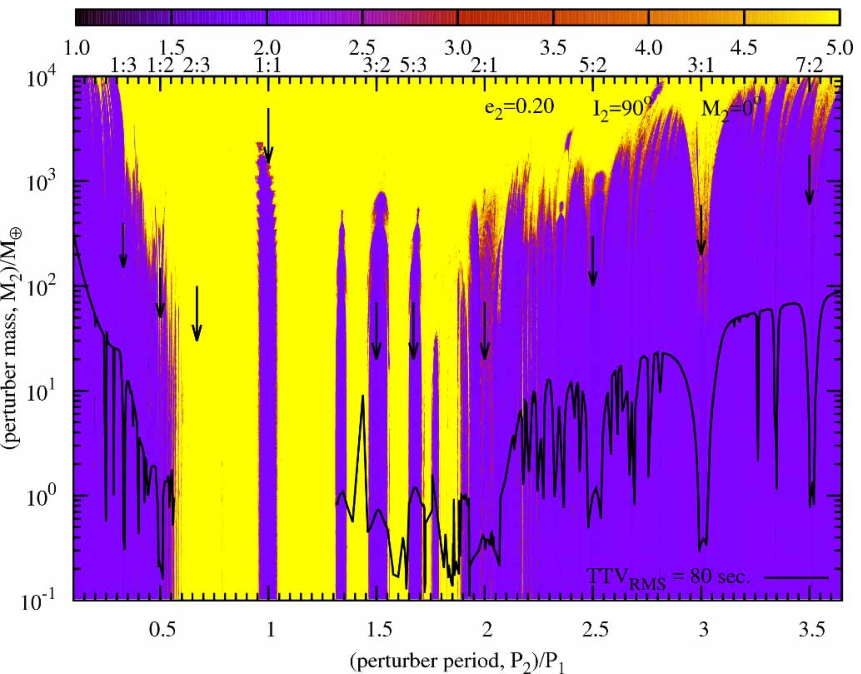}}
\caption{MEGNO stability maps superimposed on an upper mass-limit function of a hypothetical perturber. A MEGNO of around 2
indicates quasi-periodic (regular) orbits. A MEGNO of 5 or larger indicates strongly chaotic (unstable) behavior. The colors
refer to final MEGNO values after each grid integration, see text for details. For larger eccentricities of the perturber, the region around the transiting planet becomes increasingly chaotic (unstable). Vertical arrows indicate the location of
$(P_2/P_1)$ orbital resonances between the transiting planet and a perturber. The black line represents mass-period
parameters for which the perturber introduces a TTV signal with a root-mean-square of 80 seconds as determined by our timing analysis. Each panel explores a survey in different initial eccentricity of the perturber ranging
from circular to 0.20. In all maps the pair $(\omega_2, \Omega_2)$ was set to zero initially. \emph{A color version of this
figure is available in the electronic version of this manuscript}.}
\label{megnottv1}
\end{figure*}

\begin{figure*}
%Toby's internal bookkeeping notes:
%IDL script to generate each plot:
%Path (on kapc):
\centering
{\includegraphics[width=0.49\textwidth]{f5_1.eps}}
{\includegraphics[width=0.49\textwidth]{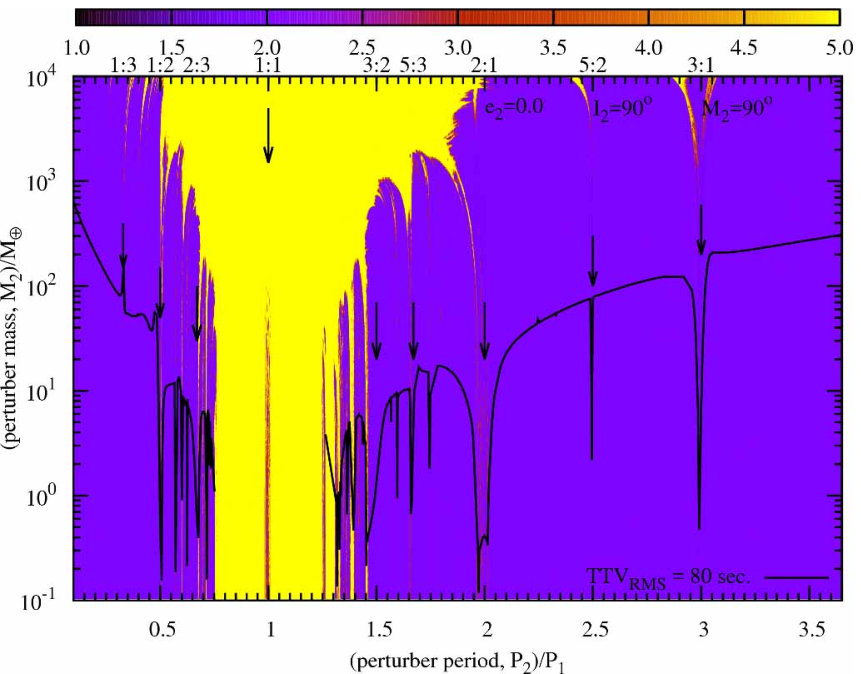}}
{\includegraphics[width=0.49\textwidth]{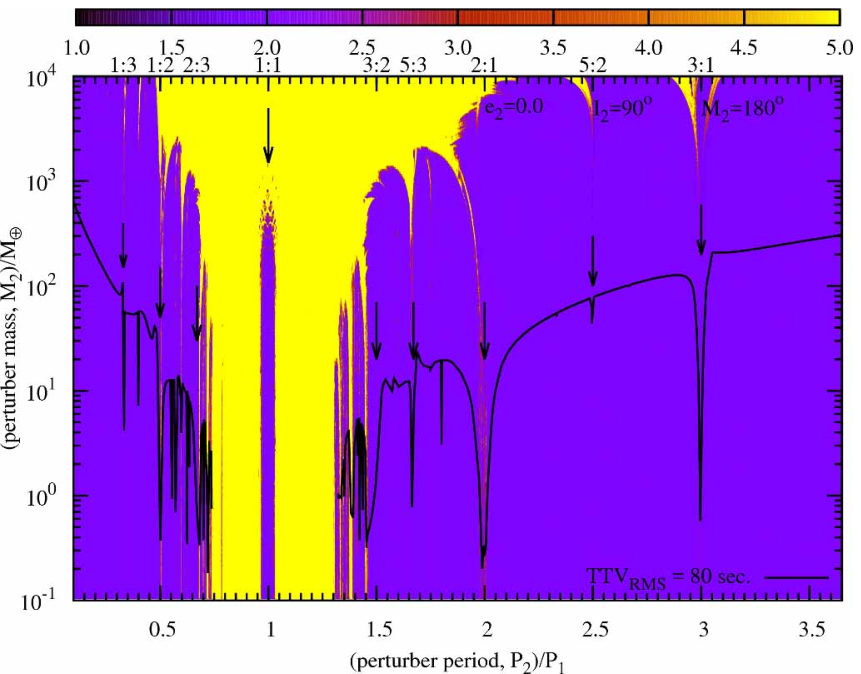}}
{\includegraphics[width=0.49\textwidth]{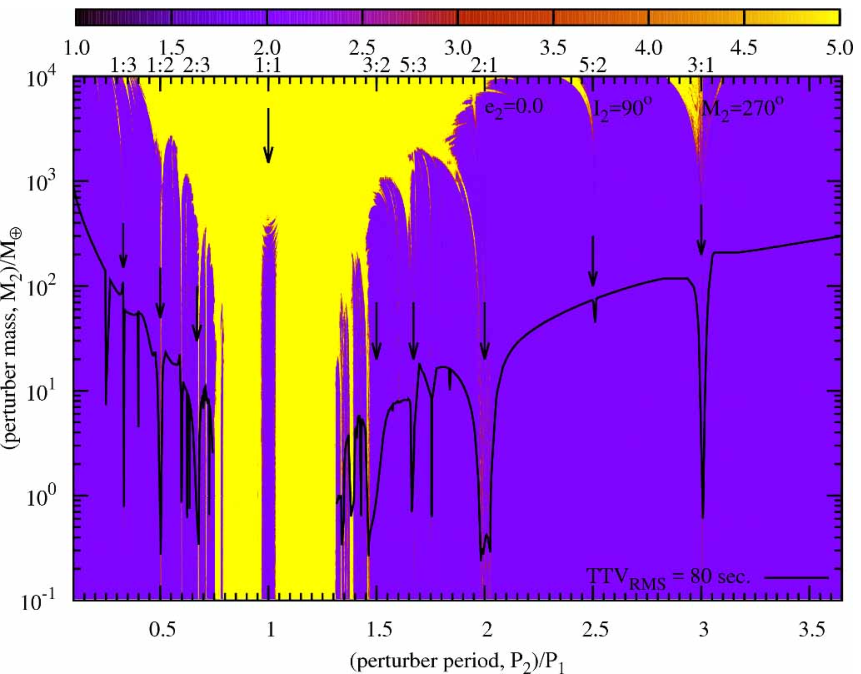}}
\caption{Same as fig. \ref{megnottv1}, but this time exploring different initial mean anomalies of the perturbing planet. Qualitatively, the quasi-periodic central co-orbital 1:1 resonance now ceases significance and is replaced by general chaos (mainly for $M_2 = 90^{\circ}$). \emph{A color version of this figure is available in the electronic version of this manuscript}.}
\label{megnottv2}
\end{figure*}

\begin{figure*}
%Toby's internal bookkeeping notes:
%IDL script to generate each plot:
%Path (on kapc):
\centering
{\includegraphics[width=0.49\textwidth]{f5_1.eps}}
{\includegraphics[width=0.49\textwidth]{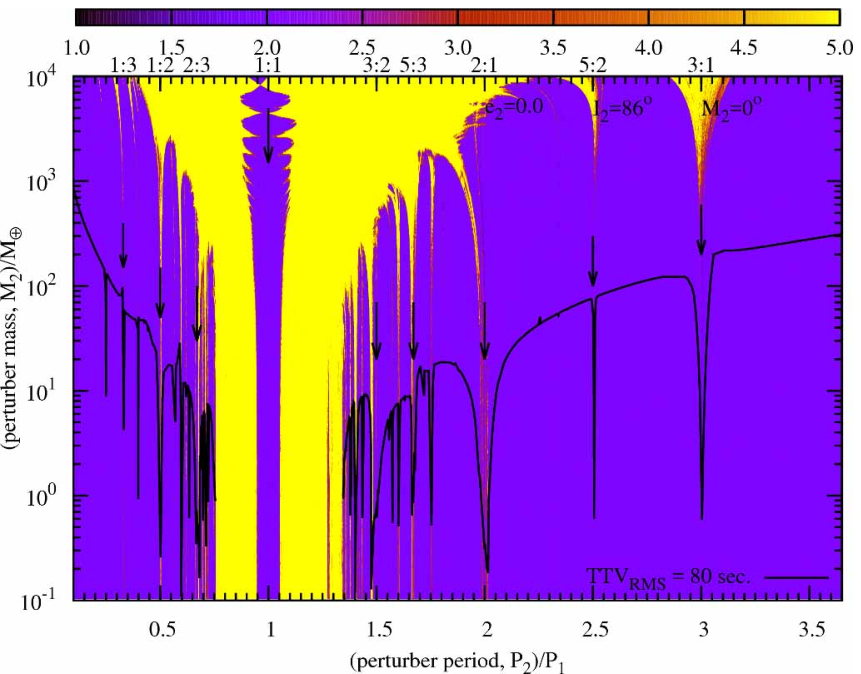}}
{\includegraphics[width=0.49\textwidth]{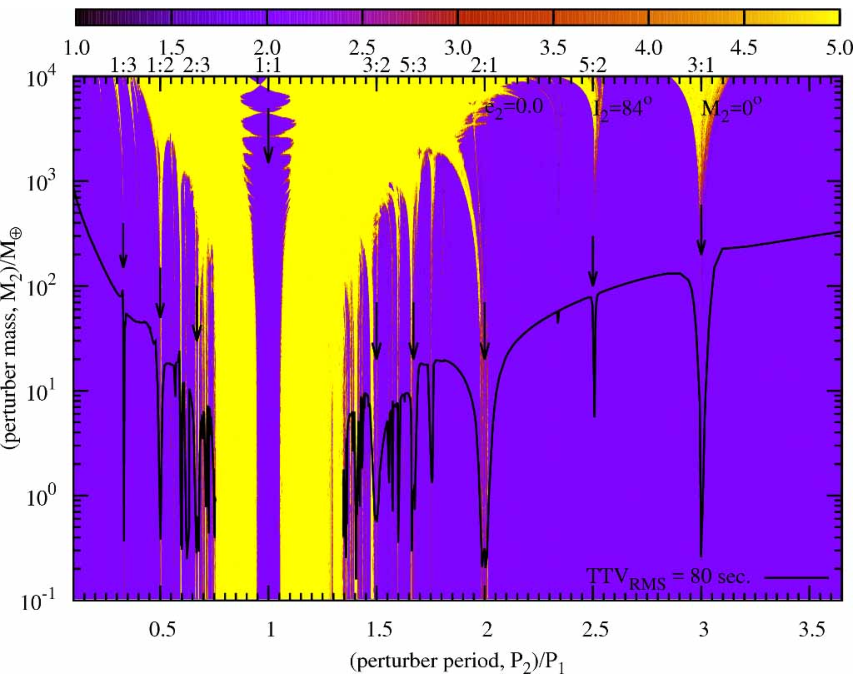}}
{\includegraphics[width=0.49\textwidth]{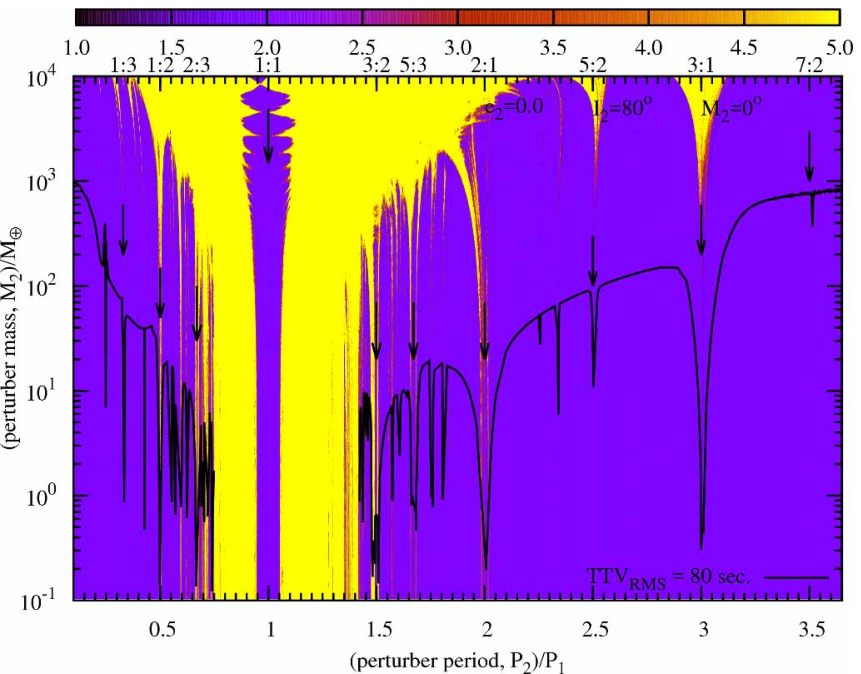}}
\caption{Same as fig. \ref{megnottv1}, but this time exploring different initial orbital inclinations of the perturbing planet. Qualitatively, the effect of varying the perturbing planet's inclination is small. \emph{A color version of this figure is available in the electronic version of this manuscript}.}
\label{megnottv3}
\end{figure*}

Our results are shown in Figure~\ref{megnottv1} to Figure~\ref{megnottv3}. We explored the range $0.1<P_2/P_1<3.65$ in orbital period ratio by fixing the osculating elements of the transiting planet to its best-fit values and allowing the perturbing planet to start in the interval from 0.02 to 0.10 AU. The mass range of the perturber was probed from 0.1 to 1000 $M_\earth$. In the figures, we show the resulting MEGNO with the perturber mass-limit function superimposed (corresponding to the 80 seconds as obtained from our timing analysis). Timing variation signals with a larger (smaller) RMS scatter will be shifted upward (downward) \citep{hoyer2011}. We mark the location of orbital resonances by vertical arrows. Furthermore, in each panel, we study the effect of various initial conditions of the perturbing planet on the overall dynamics and the resulting mass-limit function. In general, we see that the system becomes unstable/chaotic for configurations where the two planets are in close proximity to each other, and where the interactions are thus strongest. The calculation of the RMS statistic proves more difficult when the two bodies interact strongly, which can generate inclination perturbations that destroy transit regularity. As a consequence, for such cases, numerical data is sparse.

Considering exterior orbits, we can conclude that the upper mass-limit of a perturber in the 3:1 orbital resonance is around $0.5~M_{\earth}$, and for the 2:1 resonance, the limit is somewhat smaller at around $0.3~M_{\earth}$. The mass-limit for a perturber in the 7:2 or 5:2 resonance is strongly dependent on the initial conditions. For interior orbits, a perturber of a maximum of $0.5~M_{\earth}$ in the 1:2 resonance would produce a TTV signal with a root-mean-square of 80s for nearly all initial conditions. The upper mass-limit at the 3:1 resonance depends on the initial conditions, as can be seen from the figures. In the case of co-planar orbits and sufficient photometric precision, however, an interior perturber would potentially have revealed itself by a transit signal. In between all low-order resonances (exterior as well as interior) we detect higher-order resonances at which one could potentially detect the presence of a relatively low mass planets.

%old text:
%including%See \cite{2008ApJ...684...18C}%, then see \citet{2008ApJ...684...18C}, finally see \citep{2008ApJ...684...18C}
%\subsection{Gravitational TTV}
%1). how can we get the constraints on the additional planet parameters based on gravitational TTV.
%Figure 5 (the constraints on additional planet perturber from gravitational TTV)
%In order to constraint the mass and the period of potential additional companion,\cite{2010A&A...521A..60M}

\setlength{\tabcolsep}{1.1pt}
\begin{deluxetable}{lcccccc}
\tabletypesize{\scriptsize}
\tablewidth{0pt}
\tablecaption{Mid-transit times for HAT-P25b as determined from JKTEBOP (see sec.~\ref{Lightcurevanalysis})\label{midtimes}}
\tablehead{\colhead{Epoch Number} & \colhead{T$_{C}$} & \colhead{$\sigma$}& \colhead{O-C}&\\\colhead{} & \colhead{($ \mathrm{BJD}_\mathrm{TDB}$)} & \colhead{(second)}& \colhead{(second)}}
\startdata
-351 &  2455136.671678   & 68.88  &    2.70         \\
-334 &  2455198.769490   & 32.83  &   -2.08         \\
39   &  2456561.268725   & 57.43  &   -90.84        \\
45   &  2456583.187731   & 90.75  &    91.60        \\
54   &  2456616.062365   & 91.74  &    30.50        \\
57   &  2456627.021282   & 50.77  &    71.33        \\
175  &  2457058.050613   & 103.43 &   -181.80       \\
270  &  2457405.069613   & 108.43 &   -51.25        \\
273  &  2457416.029497   & 154.05 &    72.91        \\
350  &  2457697.295638   & 56.30  &    15.10        \\
\enddata

\end{deluxetable}

\begin{deluxetable*}{lccccccc}
\center
\tablecaption{\label{masslimits}Limits on Nearby Stars for HAT-P-25}
\tablehead{\colhead{band} &  \multicolumn{5}{c}{Limiting for Annulus Centered at...} & \colhead{Ref} \\ \colhead{  }&  \colhead{ 74 AU}& \colhead{148 AU}& \colhead{297 AU}& \colhead{597 AU}& \colhead{1188 AU}}
\startdata
$Ks$ ($\Delta$mag\tablenotemark{a})             &  ...    & 1.88   & 4.27     & 6.79  & 7.78    & \citet{adams2013} \\
$Sloan z$ ($\Delta$mag\tablenotemark{a})      &  3.80   &   4.27 & 5.57     & 5.80  & ...     & \citet{wollert2015} \\
upper mass-limits\tablenotemark{b} ($\mathrm{M_{J}}$) &  ...    & 568.22   & 201.98     & 69.41  & 27.37    &   \\
\enddata
\tablenotetext{a}{E.g., an object with $\Delta$mag = 5 near a star with magnitude = 12 is assigned a magnitude of 17.  }
\tablenotetext{b}{Based on $Ks$ magnitude, we calculated the upper mass-limits of nearby stars at the specified distance. }
\end{deluxetable*}

\subsection{The light travel time effect and the mass-limit of a long-term perturber}

%TTV analysis provides a powerful approach of exploring extra-solar planets, but the gravitational perturbation from the perturbing planet is not the only source of TTV. The gravitational interaction between the host star and perturbers can also give rise to TTV. The significant gravitational interaction between planet and planet is the main TTV source, when they are close proximity, even at orbital resonances. But, as the distance between planet and perturber getting longer, the movement of host star caused by the gravitational interaction between it and long-term perturber becomes the main source of TTV. The periodic mid-eclipse time variation from this movement of host star, which called the light travel time effect (LTTE; \citealt{1952ApJ...116..211I}; \citealt{1959AJ.....64..149I}). And this kind of TTV ($\mathrm{TTV_{LTTE}}$) can be expressed by the equation as follow,

In \S 4.3, we discussed the TTVs caused by the gravitational interactions between the transiting planet and a perturber. This, however, is not the only source of TTVs. The interaction between planet and perturber becomes weaker when the distance between them becomes longer. For large period ratios, the movement of the host stellar barycenter caused by the perturber becomes the dominant source of TTVs. This is known as the light travel time effect ($\mathrm{TTV_{LTTE}}$; \citealt{1952ApJ...116..211I}; \citealt{1959AJ.....64..149I}), and is calculated  by:

\begin{equation}
                   \mathrm{TTV_{LTTE}}=\case { {(G)}^{1/3} {P}^{2/3} (1-e^2) m_p \mathrm{sin}(i)} {c{(2 \pi) }^{2/3}{(m_{*}+m_P)}^{2/3}} \case{\mathrm{sin}(\omega +f[t])}{1+e\mathrm{cos}(f[t])}
\label{TTV}
\end{equation}
where the true anomaly, $f$, is the function of time, $\omega$ is the argument of the periastron, $e$ is the eccentricity, $i$ is the inclination, and $m_{*}$ is the mass of the HAT-P-25 host star, $m_{P}$ is the mass of the perturber, $c$ is the speed of light, and $G$ is the gravitational constant.

\begin{figure}
\begin{center}
\includegraphics[width=0.5\textwidth]{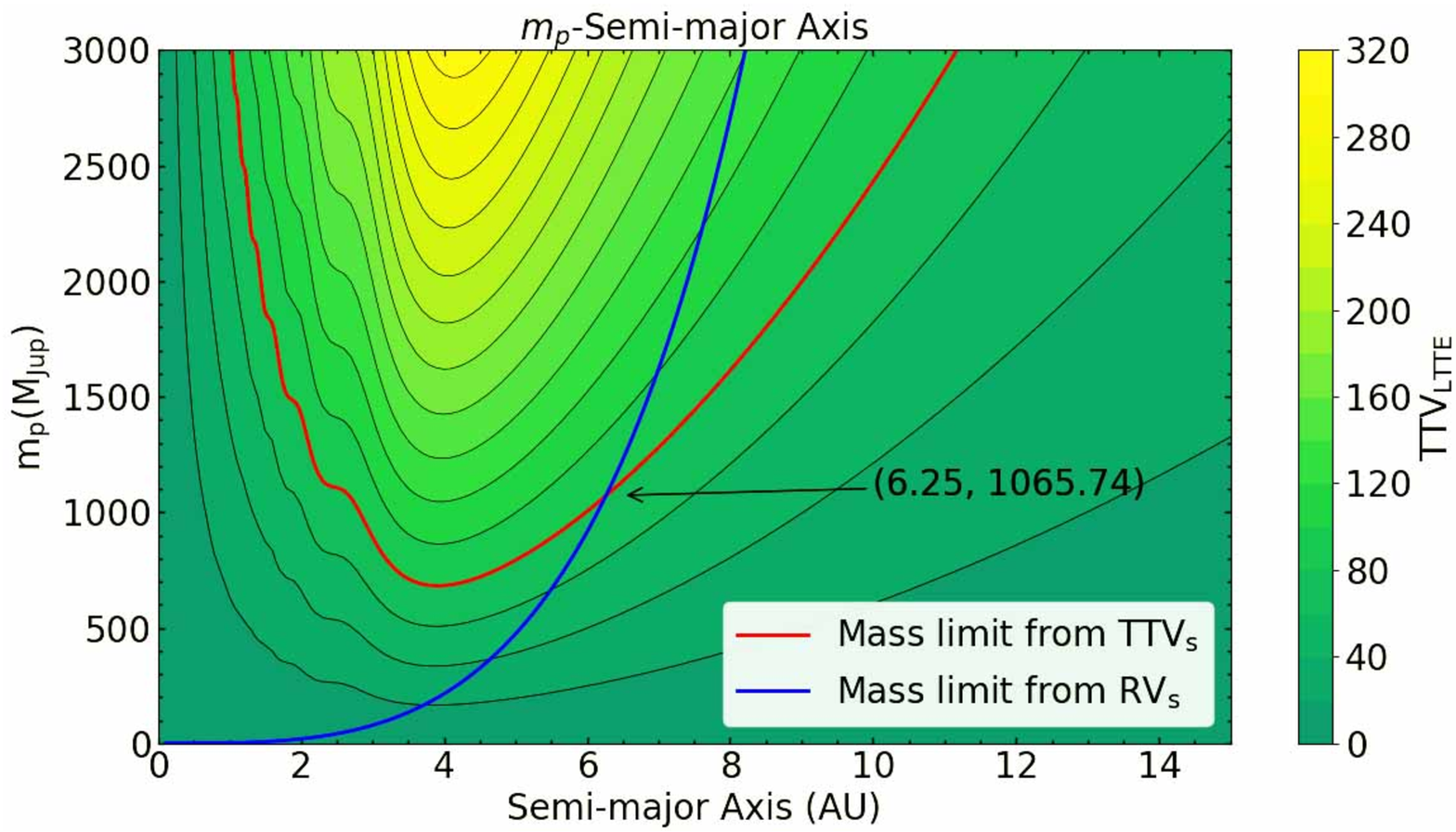}
\caption{\label{LTTE}Contours of RMSs of TTVs, the curve of the TTVs' RMS of 80s and the RMS of 3.86s curve of the residual of the best-fit RVs, with RMS in color scale. The black solid contours mark the value of the RMS across the $m_{p}$-Semi-major Axis space. The red curve and the blue curve represent the upper mass-limit derived from the analysis of TTVs and the analysis of RVs respectively. }
%The slope of blue dash line results for a mis-determined period.
\end{center}
\end{figure}

As described in \S 4.2, the TTVs we observed do not show significant timing variations, and have an RMS of only $\sim$80s. This limits the upper mass of the hypothetical perturber according to  $\mathrm{TTV_{LTTE}}$ \citep{2010A&A...521A..60M}. We calculated this upper mass for every hypothetical semi-major axis and TTV RMS. We then created a heat map as shown in Figure~\ref{LTTE}. The curve of RMS of $\mathrm{TTV_{LTTE}}$ =80s is highlighted in red. Planets of lower mass than delimited by this curve are not excluded from the system by our observations.

For our calculation, we assumed that the orbits of the transiting planet and the perturber are circular and coplanar,  implying that $e$ is 0, $\omega$ is 0, and $i$ is $90^{\circ}$, thereby giving a conservative estimate of the upper mass-limit of long-term perturbers.

In parallel, the lack of a significant trend in the RVs (e.g., \citealt{2007ApJ...657..533W}; \citealt{2014ApJ...785..126K}) provides an opportunity to determine an independent upper mass-limit on hypothetical perturbers through the radial velocity reflex velocity relation
\begin{equation}
    \mathrm{RV}=K[\mathrm{cos}(f[t]+\omega)+e\mathrm{cos}\omega]
    \label{RV1},
\end{equation}
where K is the radial velocity semi-amplitude which is calculated by
\begin{equation}
    K=(\case{2\pi G}{P(m_{*}+m_P)^2})^{1/3} \case{m_2\mathrm{sin}i}{\sqrt {1-e^2}}.
\label{RV2}
\end{equation}
The curve of the upper mass-limit from the analysis of RVs is highlighted in blue in the Figure~\ref{LTTE}.

According to our calculations, the limit from the analysis of RVs is more sensitive when the semi-major axis is short (shorter than 6.25 AU in this case). On the contrary, the limit from the analysis of TTVs is effective as the semi-major axis exceeds 6.25 AU. Synthesizing the analysis of the RMS of TTVs and the RVs, the upper mass-limit of perturbing stellar companion in the distance of 0 to 11.17 AU is 3000 $\mathrm{M_{J}}$.

%\uline{Moreover, two reported observations} \citep{adams2013, wollert2015} \uline{of high spatial resolution for this system make it possible to further constrain the magnitude of long-term star perturber. The magnitude limits of long-term perturbers for annuli at distances of $0.25\arcsec$-$4.0\arcsec$ are shown in Table~\ref{limitsmag}.}

Moreover, there are two observations \citep{adams2013, wollert2015} of high spatial resolution for this system. In these works, they put the magnitude limits on the nearby stars around HAT-P-25. Hence, we can calculate the mass-limits on the nearby stars according to the relation of absolute magnitude in $Ks$ and stellar mass \citep{pecaut2013}. Based on the magnitude limits given in \citet{adams2013} and \citet{ wollert2015}, we obtained the upper mass-limits for HAT-P-25's nearby stars by linearly interpolating the mass-magnitude relation given by \citet{pecaut2013}. The magnitude limits and upper mass-limits are listed in Table~\ref{masslimits}.

\section{Discussion and Conclusion}
Between Nov 2013 and Nov 2016, we obtained eight transit light curves of the hot Jupiter HAT-P-25b, which quintuples the number of literature transits available for this system to date.
Based on the analysis of our new photometric data, along with two follow-up light curves and RV data obtained from discovery work \citep{2012ApJ...745...80Q}, we presented new estimates of the system parameters for HAT-P-25, which is consistent with those in discovery work. Moreover, we significantly improved linear ephemeris (T$_{C[0]}$=2456418.80996$\pm$0.00025 [$\mathrm{BJD}_\mathrm{TDB}$]
 and P=3.65281572$\pm$0.00000095 days); our improved orbital period is 1.78s shorter than previous one reported in discovery work.

The analysis of $\mathrm{TTV_{G}}$ allowed us to place an upper mass-limit of a  hypothetical nearby perturbing planet as a function of its orbital separation. Near the 1:2, 2:1, and 3:1 resonances with HAT-P-25b, a perturber with mass greater than 0.5, 0.3 and 0.5 $\mathrm{M_{\oplus}}$ can be excluded, respectively. The mass-limit for a perturber in the 1:3, 5:2, and 7:2 resonance are strongly dependent on the initial conditions.
Moreover, the analysis of $\mathrm{TTV_{LTTE}}$ allowed us to present the upper mass-limit of the long-term perturber. A long-term perturber with mass greater than 3000 $\mathrm{M_{J}}$ within 11 AU of the star can be excluded.

One might struggle to counter the argument that the results of the curent paper are mundane. Our analysis reinforces the conclusion that the HAT-P-25 system fits into the well-worn hot Jupiter narrative \citep{Winn2015}. A short-period Jupiter-like planet, bereft of any detectable companions, orbits an otherwise unremarkable sun-like star at a distance far inside the radius at which the core accretion process \citep{Pollack1996} is conventionally held to operate.

Our view is that follow-up observations of the type reported here are important. Some of the most illuminating insights gained from the exoplanets have flowed from the rare hot Jupiters that turn out to have companions. For example, the outer worlds orbiting Upsilon Andromedae \citep{Butler1999} provided the first opportunity to investigate the effects of large-scale, potentially dissipation-free orbital instability \citep{Ford2005}. The HAT-P-13 system \citep{Bakos2009} produced an entirely unanticipated invitation to probe the interior structure of a planet beyond the solar system \citep{Batygin2009, Buhler2016}. More recently, the WASP-47 system \citep{Hellier2012}, with its paradigm-confounding architecture \citep{Becker2015} has offered an abundance of clues to how planetary systems form and evolve.

In the context of the forthcoming NASA's TESS and ESO's GAIA mission, more meaningful researches can be conducted with TEMP: 1).  More precise estimate of a upper mass of long-term perturber can be given through the mutual verification of the $\mathrm{TTV_{LTTE}}$ and the astrometric measurement performed by GAIA. 2). The masses of nearby perturbers (especially near resonance perturbers), however, can only be effectively constrained by the $\mathrm{TTV_{G}}$,  because astrometric technique is not sensitive to this type of perturbers. 3). Most importantly, time baseline for typical TESS field is only about 27 days while the period of typical $\mathrm{TTV_{G}}$ signal is several years. Hence, the high-precision photometric follow-up observations are required to perform decisive TTV analysis.

Over the past decade, a number of best practices for ground-based transit follow-up have emerged, spanning differential photometry, precise time handling, and Bayesian model slection. As outlined in this paper, we have incorporated these developments into our procedures. Given the unique longitude of our facilities, and our consistent access to meter+ class telescopes, we are confident that our systematic approach will allow us to eventually add to the list of landmark multiple-planet systems that contain hot Jupiters.

\clearpage

\section{Acknowledgement}
S.W. thanks the Heising-Simons Foundation for their generous support.

This research was supported by
the National Natural Science Foundation of China and the Chinese Academy of Sciences Joint Fund on Astronomy (No. U1431105);
the Key Development Program of Basic Research of China (No. 2013CB834900);
the National Basic Research Program of China (Nos. 2014CB845704, and 2013CB834902); the National Natural Science Foundation of China (No. 11333002, 11373033, 11433005, 11673027, 11403107, 11503009, 11673011);
the Young Scholars Program of Shandong University, Weihai (No. 2016WHWLJH07);
the Natural Science Foundation of Jiangsu Province (BK20141045);
the National Defense Science and Engineering Bureau civil spaceflight advanced research project (D030201);
the KASI grant 2016-1-832-01 and 2017-1-830-03.
Results from numerical simulations were partly performed by using a high performance computing cluster at the Korea Astronomy and Space Science Institute.

\clearpage
%\bibliography{reference}

\end{document}